\documentclass[preprint2]{aastex}
\newcommand{\kms}{km s$^{-1}\;$}
\newcommand{\kmss}{km s$^{-1}$}

\newcommand{\vlsr}{$V$$_{\rm LSR}$}

\newcommand{\lsun}{\mbox{$L$$_{\sun}$}}

\newcommand{\ho}{H$_{2}$O$\;$}

\newcommand{\mb}{mJy beam$^{-1}$}

\shorttitle{Water masers in NGC\,6240 and M51}
\shortauthors{Hagiwara and Edwards}
\begin{document}
%
%
\title{High-Resolution Imaging of Water Maser Emission in the active galaxies NGC 6240 and M51}
%
\author{Yoshiaki Hagiwara\altaffilmark{1} and Philip G. Edwards\altaffilmark{2}}
\affil{$^{1}$Natural Science Laboratory, Toyo University, 5-28-20, Hakusan, Bunkyo-ku, Tokyo 112-8606, Japan; yhagiwara@toyo.jp} 
%
%
\affil{$^{2}$CSIRO Astronomy and Space Science, P.O. Box 76, Epping NSW 1710, Australia; Philip.Edwards@csiro.au}
%
%
\affil{Received 2015 September 23; accepted 2015 November 10; published 2015 December 17}
\begin{abstract}
We present the results of observations of 22 GHz H$_2$O maser emission
in NGC\,6240 and M51 made with {the Karl G.\ Jansky Very Large Array}.
%
%
Two major H$_2$O maser features and several minor features are detected toward
the southern nucleus of NGC\,6240. These features are
redshifted by about 300 \kms from the galaxy's systemic velocity and
remain unresolved at the synthesized beam size.
A combination of our two-epoch observations and published data
reveals an apparent correlation between the strength of the
maser and the 22 GHz radio continuum emission, implying that the maser
excitation relates to the activity of an active galactic nucleus in
the southern nucleus rather than star-forming activity.  The
star-forming galaxy M51 hosts H$_2$O maser emission in the center of the
galaxy; however, the origin of the maser has been an open question.  
We report the first detection of 22 GHz nuclear radio continuum emission
in M51. The continuum emission is co-located with the maser position, 
which indicates that the maser arises
from active galactic nucleus-activity and not from star-forming activity in the
galaxy.

%
\end{abstract}
%
\keywords{galaxies: active - galaxies: individual (NGC6240, M51) - galaxies: ISM - galaxies: nuclei - masers}
\section{Introduction}

Observations of 22 GHz \ho masers have become established as a powerful
tool for investigating the inner few parsecs of active galactic nuclei
(AGNs), and have enabled precise distance measurements to local group galaxies 
and nearby AGNs \citep[e.g.,][]{herr99,linc09}. 
Observations of extragalactic masers are of interest for determining
the origin of the maser emission (e.g., whether it is driven by AGNs or starburst 
activity), and the suitability of the galaxy for direct distance 
measurement.
Extragalactic \ho masers have also been {detected} in the sub-millimeter bands,
with the 183\,GHz, and possibly 439\,GHz, transitions detected in NGC\,3079 \citep{liz05}, 
and the 321\,GHz transition detected in the Circinus galaxy \citep{hagi13}, 
providing additional information on the physical environments 
hosting these masers.
In this paper we describe the results of observations 
{with the Karl G.\ Jansky Very Large Array (VLA)} of
lower-luminosity 22\,GHz \ho maser emission in the galaxies 
NGC\,6240 (J1652+0024) and M51 (J1329+4713).

{\object{NGC\,6240} is an interacting/merging galaxy hosting 
two nuclei that are separated by $\sim$1.$\arcsec$5} \citep{bes01}. At 22\,GHz, an \ho maser with low isotropic luminosity 
(L$_{\rm {H_2O}}$  $\la$ 10 $\lsun$) 
has been reported in this galaxy \citep{hen84,hagi02,naka02}.
NGC\,6240 has a large far-infrared luminosity $\approx$ L$_{\rm FIR}$ = 10$^{11-12}$ \lsun, 
which cannot be
accounted for by starburst activity alone, but which can be more naturally
explained as a buried AGN heating surrounding dusty components
\citep[e.g.,][]{dep86,sand88,ris06}.
Neutral Fe K$\alpha$ lines at 6.4\,keV due to reflection from optically
thick material are detected from both nuclei, which is clear 
evidence for both being AGNs
\citep{kom03}.  Very Long Baseline Interferometry (VLBI) observations
\citep{gall04,hagi11} detected compact radio sources with brightness
temperatures of 7$\times$10$^6$\,K for the northern nucleus and 
1.8$\times$10$^7$\,K for the southern nucleus 
\cite[e.g.,][]{col94}, confirming the presence of AGNs at
centimeter radio wavelengths.  
{Using the VLA, we found that the maser emission was coincident
with the southern nucleus to within the 0.$\arcsec$1 resolution of
the observation} \citep{hagi03}. The origin of the maser has been an
open question for some time \citep{hagi10}.

The star-forming galaxy \object{M51} (\object{NGC\,5194}) hosts a Seyfert/LINER nucleus \citep{heck80}
and has \ho maser emission \citep{ho87} that consists of Doppler-shifted features on
either side of the galaxy's systemic velocity \citep{hagi01}. 
The redshifted features
are located toward the nuclear radio source, while the
location of the blue-shifted counterpart is displaced from the radio
source toward the southeast.
A small velocity gradient closely aligned with the radio jet
is detected from the redshifted features \citep{hagi07}. The redshifted maser most
likely amplifies the background radio continuum jet, while the
blue-shifted counterpart marks off-nuclear star formation in the
galaxy.

{The upgraded VLA capabilities},
with bandwidths of 2 GHz and improved spectral
resolution \citep{perl11}, have allowed us to address the nature of the 
low-luminosity \ho masers in \object{NGC\,6240} and \object{M51}
and to map the distribution of maser spots on the 0.1 arcsec scale.

We adopt cosmological parameters of H$_{0}$ = 73 \kms Mpc$^{-1}$,
$\Omega$$_{\Lambda}$ = 0.73, and $\Omega$$_{M}$ = 0.27. 
The luminosity distance to \object{NGC\,6240}
is therefore 103 Mpc (z=0.02448) and 10.6 Mpc (z=0.002) to
\object{M51}.  Thus, 1 arcsec corresponds to a linear scale of
475\,pc for NGC6240 and 51\,pc for M51.
We adopt the optical velocity definition
and velocities are calculated with respect to the local standard
of rest (LSR).

\section{OBSERVATIONS AND DATA REDUCTION}
We observed NGC\,6240 with the  National Radio
Astronomy Observatory (NRAO)\footnote{The NRAO is operated by Associated Universities,
Inc., under a cooperative agreement with the National Science
  Foundation.} {VLA} in the B configuration on
2012 July 21 (epoch 1) and 2012 August 28 (epoch 2) and observed M\,51
in the A configuration on 2012 November 17.  Observations were made of 
22\,GHz \ho maser emission in the 6$_{16}$$-$5$_{23}$ transition, employing 16
intermediate frequencies (IFs) of 16\,MHz bandwidth with dual polarization, resulting in a total
bandwidth of 256\,MHz for each polarization. The total velocity coverage is
$\sim$3450\,\kmss, however, the velocity coverages were not continuous
in both observations: there is a $\sim$430\,MHz frequency gap
between IFs 1--8 and IFs 9--16, where IF4 is centered on the
systemic velocity of each galaxy. The total observing time for NGC\,6240 
at each epoch was $\sim$2.5 hr and that of M51 was $\sim$3 hr.
{In the NGC\,6240 observations, the phase-referencing observations were executed 
in a sequence of 3 minute scans with 2 minutes on NGC\,6240 and 40 s on J1658+0741.
For the M51 observation, the phase-referencing was performed 
in a sequence of 3.75 minute scans with 2 minutes on M51 and 40 s on J1419+5423.}
Flux density and bandpass calibration were performed using
observations of 3C\,286. Standard VLA data calibration and editing
were made using the NRAO Astronomical Image Processing Software
(AIPS). Uncertainties of flux density were estimated in the same method as 
explained in \citet{hagi10}.

After the phase and flux calibrations and data editing, the
continuum emission was subtracted from the spectral-line visibilities
using line-free channels prior to the imaging and CLEAN deconvolution
of the maser emission. Imaging and CLEANing of the continuum were done
separately.

The synthesized beam sizes produced from naturally weighted
spectral-line images were 0.38$\times$0.34 arcsec$^2$ (PA=18$\degr$)
for NGC\,6240 and 0.10$\times$0.09 arcsec$^2$
(PA=$-$3$\degr$) for M51.  The spectral resolution was 125\,kHz
(1.7\,\kmss) per spectral channel for all observations.  The rms noise
levels of the naturally weighted images at 125 kHz resolution were 
$\sim$1.5$-$2.0\,\mb (epoch 1), $\sim$0.8$-$1.3\,\mb (epoch 2) for NGC\,6240 and 
$\sim$1.0$-$1.2\,\mb\,for M51.  The rms noises of the continuum images were 
about 0.14\,\mb\, for NGC\,6240, and 0.036\,\mb\, for M51, respectively.


\section{RESULTS}
\subsection{NGC\,6240}
Figure~\ref{figure1} shows two-epoch spectra of the \ho maser emission
from the southern nucleus in the velocity range spanning 
\vlsr=7450$-$7690 \kmss. The total velocity coverages in our observations
are \vlsr $\approx$ $-$730$-$990 \kms and $\approx$ 6590 -- 8320
\kmss. 
The maser line peaking at \vlsr=7609.3 \kms is clearly detected at
both epochs, though it is somewhat weaker at the second epoch.
This location of this feature 
was determined with the VLA in 2012 to 
be coincident with the southern nucleus.
A marginally significant feature at \vlsr=7563.2 \kms in the first epoch
matches features 
seen by single-dish measurements in 2000$-$2001, and 2005 (see Table 2),
but there is no evidence of this feature in the second epoch. 

%
Figure~\ref{figure2} shows the spectra of tentative features peaking at
\vlsr=7158\kms, 7372 \kms, and 7396 \kms and in the southern nucleus.
The maser emission also exhibits a broad profile that ranges from \vlsr $\approx$ 7360 to 7400 \kmss.
These features are detected, tentatively, for the first time in the
galaxy.  No other \ho maser is detected at either epoch 
above the 3$\sigma$ level of $\sim$3\,\mb\, in the 
spectra from the northern nucleus, or at any other point 
within a field of view 7.7$\arcsec$
$\times$ 7.7$\arcsec$, centered on the southern nucleus. 

Figure~\ref{figure3} shows the 22\,GHz continuum image from epoch 1
data, obtained by integrating over the full frequency range {after the removal of the maser line emission}. The image
shows two major radio sources, the southern and northern nucleus, both
coinciding with the radio sources at lower frequencies with similar
angular resolution \citep[e.g.,][]{car90, col94, bes01}. Extended
components seen in lower frequency VLA images in the literature
\citep[e.g.,][]{col94, baa07} were resolved out in our continuum
image. The flux densities of the 22\,GHz continuum of the two nuclei and
\ho maser are summarized in Table~\ref{table1}.

All of the masers detected in our observations remain unresolved at
the angular resolution of $\sim$ 0.$\arcsec$2, or 100\,pc. The position
of the 7609\,\kms feature is {
$\alpha$(J2000): 16$^{\rm h}$52$^{\rm m}$58$\fs$886, 
$\delta$(J2000): +02$\degr$24$\arcmin$03.$\arcsec$260, from which all other
features reside within $\approx$ 0.1$\arcsec$, about a half of the synthesized beam size of our observation. 
{Positional accuracy, without consideration of systematic errors, 
 is $\sim$ $\theta_{\rm beam}$/2$\times$S/N $\sim$ 0$\arcsec$.02, where $\theta_{\rm beam}$
is the synthesized beam of our observations and S/N is the signal-to-noise ratio of the maser spectra.
The position uncertainties considering the position error of the calibrator source \citep{reid14} are estimated to
be   $\Delta$$\alpha$ $\sim$ 0.$^s$02  and $\Delta$$\delta$ $\sim$  0.036$\arcsec$.}

The positions of these masers coincide with that of the southern
nucleus, which is consistent with results of the previous VLA
observations \citep{hagi10}. The estimation of the relative positional
error between the maser and the southern continuum peak ($\sim$3 pc or
less) is considered in \citet{hagi03} and \citet{hagi10}.  A
cumulative list of detected \ho maser features in the galaxy, which is
revised from the table~2 in \citet{hagi10}, is presented in
table~\ref{table2}.

Figure~\ref{figure4} shows the time variability of the maser and nuclear
continuum flux densities, taking three-epoch VLA data obtained from
2002--2009 (see Figure~3 in \citet{hagi10}.
There is a similar decrease, in absolute terms,
in the flux densities of the southern nucleus continuum emission and 
the water maser emission over one decade. 
With two epochs in 2012 and three earlier epochs,
the result is strongly suggestive, but further monitoring is
required before a conclusive claim for correlation can be made.

\subsection{M51}
Figure~\ref{figure5} shows the spectrum of the maser at 
\vlsr=453.4$-$668.2 \kmss. No maser has been detected outside this range,
with the full range searching covering \vlsr $\approx$ $-$400 -- 1300 \kmss.
The detected maser features are redshifted with respect to the galaxy's systemic velocity of \vlsr=472$\pm$3\kms \citep{sco83}.
All of the detected maser emission remains unresolved at the synthesized
beam size of 0.1 arcsec. The positions of the features in the
spectrum are all confined within $\approx$ 0.01$\arcsec$ from the position of a feature at 
\vlsr= 565.1 \kms: $\alpha$ (J2000) = 13$^{\rm h}$29$^{\rm m}$ 52.708$^{\rm s}$,
$\delta$ (J2000) = +47$\degr$11$\arcmin$42.810$\arcsec$,} which is
consistent with those of earlier measurements in the VLA A configuration
\citep{hagi01, hagi07}.

A naturally weighted 22\,GHz continuum image of M51 is shown in
Figure~\ref{figure5}. The continuum emission has been detected with
an S/N of $\sim$6. This is the first
reported detection of the 22\,GHz radio continuum in M51. The
image presents a nuclear source located at {
$\alpha$(J2000) = 13$^{\rm h}$29$^{\rm m}$52$\fs$708,
$\delta$(J2000) = +47$\degr$11$\arcmin$42.812$\arcsec$, where the relative errors between the
continuum source and the maser are $\sim$ 0''.01--0''.025).}  Thus, the positions of the unresolved
maser sources and continuum emission are located in the same position
in the galaxy within errors. The continuum image shows a jet component
that is nearly northeast at PA = 158 $\pm$ 19$\degr$, the axis of which
is consistent with that on the lower frequency VLA images
\citep{brad04, dum11}.
Table~\ref{table3} provides a summary of radio flux densities and the spectral index of the radio source.
In the VLA full-track observation performed in 2003 July, a weak
velocity gradient from \vlsr=554 to 563 \kms was identified at 
PA=155$\degr$, nearly along the nuclear radio jet \citep{hagi07}; however;
in our new observation no distinct velocity gradient is apparent.
This may be the result of source evolution, though we note 
that the 2012 epoch had an observing duration of 3 hr
as opposed to 8 hr in the 2003 observation, and somewhat
poorer ($u,v$) coverage. The spectral sensitivity does not differ  
significantly from the previous run conducted in 2003 \citep{hagi07}. 
\begin{deluxetable}{ccccccccc}
\tablecolumns{6}
\tablewidth{0pc}
\tablecaption{\sc  Flux Densities of the maser and double nuclei in NGC\,6240\label{table1}}
\tablehead
{
 &\multicolumn{2}{c}{Maser (\vlsr,~\kmss)}  &&  \multicolumn{5}{c}{22 GHz continuum} \\
\cline{2-3}\cline{5-9}\\
&\colhead{7563.2}&\colhead{7609.3}&&\multicolumn{2}{c}{Southern Nucleus}&& \multicolumn{2}{c}{Northern Nucleus}\\
\cline{5-6}\cline{8-9}\\
&&&&\colhead{$F_{P}$}&\colhead{$F_{T}$}&&$F_{P}$&$F_{T}$\\
(1)&(2)&(3)&&(4)&(5)&&(6)&(7) \\
}
\startdata
{2012 Jul 21~} & 6.0$\pm$1.5 & 9.7$\pm$1.5 && 9.9$\pm$0.5  & 13.4$\pm$0.7&&4.1$\pm0.2$&5.5$\pm$0.3 \\
{2012 Aug 28}  &  $<$2.7$^{\mathrm{a}}$   & 5.1$\pm$0.9 &&  7.7$\pm$0.4 & 11.9$\pm$ 0.6&&3.5$\pm$0.2&5.4$\pm$0.3
\enddata
\tablecomments{Col. (1): observing epoch; col. (2),(3): 22 GHz
  flux density (mJy) of each maser feature; col. (4)--(7): 22 GHz continuum peak
  flux (\mb) and integrated flux density (mJy) of the southern nucleus (S) and northern nucleus (N)}
  \tablenotetext{a}{3 $\sigma$ upper limit value}
\end{deluxetable}
\begin{deluxetable}{lccccc}
\tablewidth{0pt}
\tablecaption{\sc A cumulative list of \ho maser observations in NGC 6240 \label{table2}}
\tablehead{
\colhead{Epoch}           & \colhead{Telescope}      &
\colhead{Velocity range}&\colhead{Maser velocities }& \colhead{References}\\
~~~(yyyy.mm)&&(\kms, LSR)&(\kmss, LSR)&
}
\startdata
2000.03 & Green Bank & 7400$-$7700 & 7565.0$\pm$0.8& (1) \\
2001.01 & Green Bank & 6500$-$8100 & 7565.6$\pm$0.5& (1)\\
2001.05 & Effelsberg & 6850$-$7870 & 7565.0$\pm$1.1, 7609.0$\pm$1.1$^{\mathrm{a}}$&(2)\\
2001.06$^{\mathrm{b}}$& Nobeyama&6704$-$8858 & 7566.4$\pm$0.5&(3)\\
2001.12 & Green Bank & 6500$-$8100 & 7568.6$\pm$0.7& (1)\\
2002.04 & Green Bank & 6500$-$8100 & 7567.6, 7612.1$\pm$0.1& (1)\\
2002.06 & VLA & 7525$-$7665&  ~7611.0$\pm$2.6$^{\mathrm{a}}$& (4)\\
2005.01 & Nobeyama   & 6490$-$8610 & 7564.4$\pm$0.8 & (5),(6)$^{\mathrm{c}}$ \\
2007.01 & Nobeyama   & 6490$-$8610 & 7442.0, 7561.5$\pm$0.8 & (6) \\
2009.01 & VLA        & 7370$-$7660 &7491.1$\pm$0.2& (7) \\ 
2012.07 & JVLA       & ~6593$-$8321$^{\mathrm{d}}$& 7563.2, 7609.3$\pm$0.8& This article\\
2012.08 & JVLA       & ~6593$-$8321$^{\mathrm{d}}$& 7609.3$\pm$0.8&  This article\\
\enddata
{
\tablenotetext{.}{Notes.}
\tablenotetext{a}{Uncertainties of velocity substituted by the channel spacings of Effelsberg or VLA}
\tablenotetext{b}{Detected by averaging  the spectra  obtained in 2001 April and June}
\tablenotetext{c}{By averaging the spectra obtained from 2003 to 2007, a narrow line maser feature centered at \vlsr=7442.0$\pm$0.8 \kms was detected (K.Nakanishi et al. 2015, in preparation).}
\tablenotetext{d}{Velocity range of $-$734 —- 994 \kms to be added}
\tablerefs{(1)Braatz et al. (2003), (2)Hagiwara et al. (2002), (3)Nakai et al. (2002), (4)Hagiwara et al. (2003), (5)Nakanishi et al. (2008), (6)K.Nakanishi et al. (2015,  in preparation), (7)Hagiwara (2010)}
}
\end{deluxetable}
\section{DISCUSSION}
\subsection{NGC 6240}
%
From two-epoch observations in 2012 and three earlier epochs from 2002 to 2009, 
we find that the strength of
the maser and continuum emission from the AGNs in the southern nucleus
vary in a correlated fashion.
This strongly suggests that the maser in the galaxy amplifies the
background continuum emission, which is exhibiting flux variability 
that is consistent with that typically observed from the central engine in AGNs.  
These observations constitute evidence for the maser in NGC 6240
being a nuclear maser similar to the cases of other \ho megamasers
explored to date.  

Further observations at VLBI angular resolutions would
be required to determine whether the maser is a ``disk
maser,'' tracing an inclined disk located at radii $\sim$0.1--1\,pc
from the central engine \citep[e.g.,][]{linc09}.  It is also
interesting to note the tentative detections of some new weaker features 
in the second epoch, whereas the prominent features at 
\vlsr = 7563.2 and 7609.3 \kms have weakened.  
However, the emergence of the tentative features at epoch 2 is in contrast with
the fact that the 7609 \kms maser and the 22 GHz continuum emission weakened in epoch 2.

\begin{deluxetable}{lccc}
\tablewidth{0pt}
\tablecaption{\sc Radio spectra of M51 \label{table3}}
\tablehead{
\colhead{Frequency: $\nu$}  & \colhead{5 GHz$^{a}$}& \colhead{8 GHz$^{b}$} &\colhead{22 GHz}\\
}
\startdata
$S$$_{\nu}$$^{c}$ (\mb) & 0.89 & 0.21&  0.23 $\pm$ 0.04 \\
Spectral index: $\alpha_{8-22}$ (8--22 GHz) & & \colhead{0.1}  & \\
\enddata
\tablenotetext{a}{\citet{cra92}}
\tablenotetext{b}{\citet{brad04}}
\tablenotetext{c}{Peak flux densities measured by the VLA in the A-configuration}
\end{deluxetable}

It is possible
that these new features are not amplifying the background continuum but are associated 
with a starburst activity in the center of the galaxy \citep{gen98} due to 
the low-luminosity of these features, which is consistent with star-forming activity \citep{hagi07}.
Further monitoring of the velocity ranges in which these new
tentative features are seen is required to confirm their existence
and trace their variability against the nuclear continuum.

All the maser features reported previously in the galaxy were redshifted
by $\sim$ 200$-$300 \kms with respect to both the systemic velocity of
the galaxy (\vlsr=7304 \kmss) and the H$_{\rm I} $ peak velocity at N1 (7295
\kmss: the ``systemic" velocity of the southern nucleus) \citep{baa07}.
These redshifted features arise from the southern nucleus but not
from the northern nucleus. The tentative detection of the \vlsr =
7158 \kms feature, which is blue-shifted by $\sim$150 \kms from the
systemic, may constitute the first detection of counterpart to the
redshifted features.
{Although NGC\,6240 does not display the characteristic
profile of maser emission from a rotating disk (with three
distinct groups of maser features, one centered close to the systemic
velocity and the others offset on either side by up to $\sim$1000\,km\,s$^{-1}$),
we note that if the red- and blue-shifted features are interpreted as originating in an edge-on disk
\citep[see, e.g.,][]{miyo95}, the implied rotation speed ranges up to $\sim$ 300 \kmss.}
Accordingly, the possibility of
star-forming maser in NGC\,6240 is less likely despite the low
isotropic luminosity of the maser emission.

As already discussed in \citet{hagi10}, the non-detection of the maser
toward the northern nucleus is not surprising even if both nuclei
host nuclear masers, because of the very small possibility of the 
the disks around both nuclei in the galaxy being aligned
edge-on to our line of sight.

%

\subsection{M51}
%
One of the most important results in our observations is that the 
22\,GHz nuclear continuum has been detected at the location of the radio
nucleus of M51. 
This is the first detection of {the nuclear continuum source} at 
a frequency higher than 8\,GHz, due to the improved VLA
sensitivity. Moreover, we find that the location of the maser
components coincides with that of the nuclear continuum. This indicates
that the maser in M51 is a nuclear maser that is associated with
AGN activity, although the maser is classified as a ``low-luminosity maser" 
in \citet{hagi07}.  
By adopting an 8\,GHz flux density of 0.21\,mJy that was obtained by VLA A
in Table~\ref{table3}, 
the spectral index, $\alpha$ (where $S$$_{\nu}$ $\propto$ $\nu^{\alpha}$, with  
$S$$_{\nu}$ being the flux density at frequency $\nu$), 
is $\approx$ 0.10,
very flat, assuming no flux variation.
According to single-dish measurements in literature 
the total flux density is 142$\pm$15\,mJy at 22.8\,GHz \citep{kle84}. 
The total flux density of $\sim$0.4\,mJy in our 22\,GHz 
continuum map is clearly much lower,
indicating that the source is heavily resolved on these angular scales
and that we are seeing only
%
the nucleus of the galaxy. 

It is interesting to note that the jet axis that is associated with the
nuclear continuum image is consistent with that previously reported
at 5 and 8\,GHz \citep{cra92,brad04}, 
which confirms our detection and also
shows that the velocity gradient detected in the earlier VLA data is
seen along this jet axis. From our current data, there is no evidence
that the M51 maser is a disk maser surrounding the nucleus of the
galaxy at our resolution.

We failed to image the maser velocity gradient of \citet{hagi10} in
this data.  The velocity gradient was later imaged by CO (1-0) and
CO(2-1) emission at $\sim$ 0.4$\arcsec$--0.8$\arcsec$ resolution in the
nucleus of the galaxy, in good agreement with the velocity range and
gradient direction of the maser \citep{satoki07}, although these CO
molecular gases are located $\approx$ 1$\arcsec$ to the west of the
maser or the radio nucleus. The velocity gradient in the previous data
is real; however, we do not see the maser gradient in these observations, 
possibly due to the poorer $(u,v)$ coverage resulting from the shorter observing time.

The weak off-nuclear blue-shifted emission peaking at \vlsr = 445 \kms
that was detected in the earlier VLA run was not detected in our observation.
The blue-shifted maser is not in the nucleus and if the excitation of the
maser is due to star-forming activity, it would not be surprising
that the maser is weak and variable  \citep[e.g.,][]{bau96}.

%
%
%
\section{SUMMARY}
We report a correlated variation in the strength of the \ho maser and 
the 22 GHz nuclear continuum emission in the southern nucleus of
NGC\,6240. The \ho maser in the galaxy is additionally co-located 
with the nucleus and we conclude  the maser is associated
with AGN activity and not star formation.
Thus, a scenario that argues for the possibility of
a star-forming maser in the galaxy has been ruled out by our new observations.  
Several new minor H$_2$O maser features are detected at low significance toward
the southern nucleus of NGC 6240 in the second epoch. More sensitive observations are required to confirm these features.

We also report the first detection of 22 GHz nuclear radio continuum emission
in M51. The continuum emission is co-located with the maser position, 
which indicates that the maser arises
from nuclear AGN activity and not from star-forming activity in the
galaxy.

The extragalactic \ho masers in both NGC\,6240 and M51 show low
isotropic luminosity on the order of 1 \lsun, which is by two
magnitudes lower than that of the water ``megamaser," although it is most
likely that the low-luminosity \ho maser in the galaxies is a nuclear
maser. The maser most likely amplifies the background radio 
continuum emission from the host galaxy nucleus. VLBI
observations at milliarcsecond angular resolution would be able to
reveal the distribution of the maser emission.

\acknowledgments 
{We thank the anonymous referee for useful comments on our manuscript.}
This research was supported by Japan Society for the Promotion of Science (JSPS) Grant-in-Aid for Scientific Research(B)
(Grant Number: 15H03644).
This research has made extensive use of the NASA/IPAC
Extragalactic Database (NED) which is operated by the Jet Propulsion
Laboratory (JPL), California Institute of Technology, under contract
with NASA.


\begin{onecolumn}
\begin{figure}
\epsscale{1.0}\plottwo{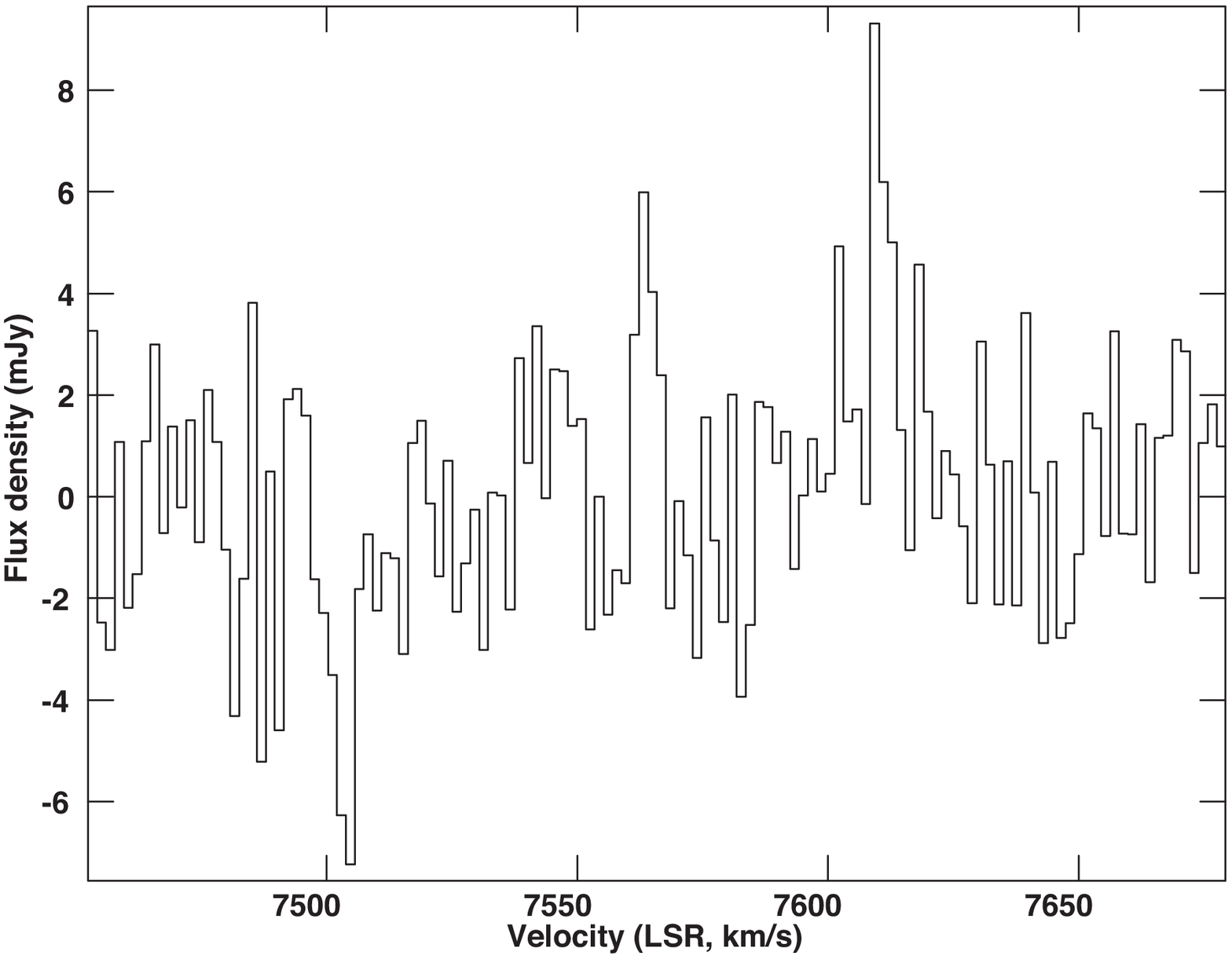}{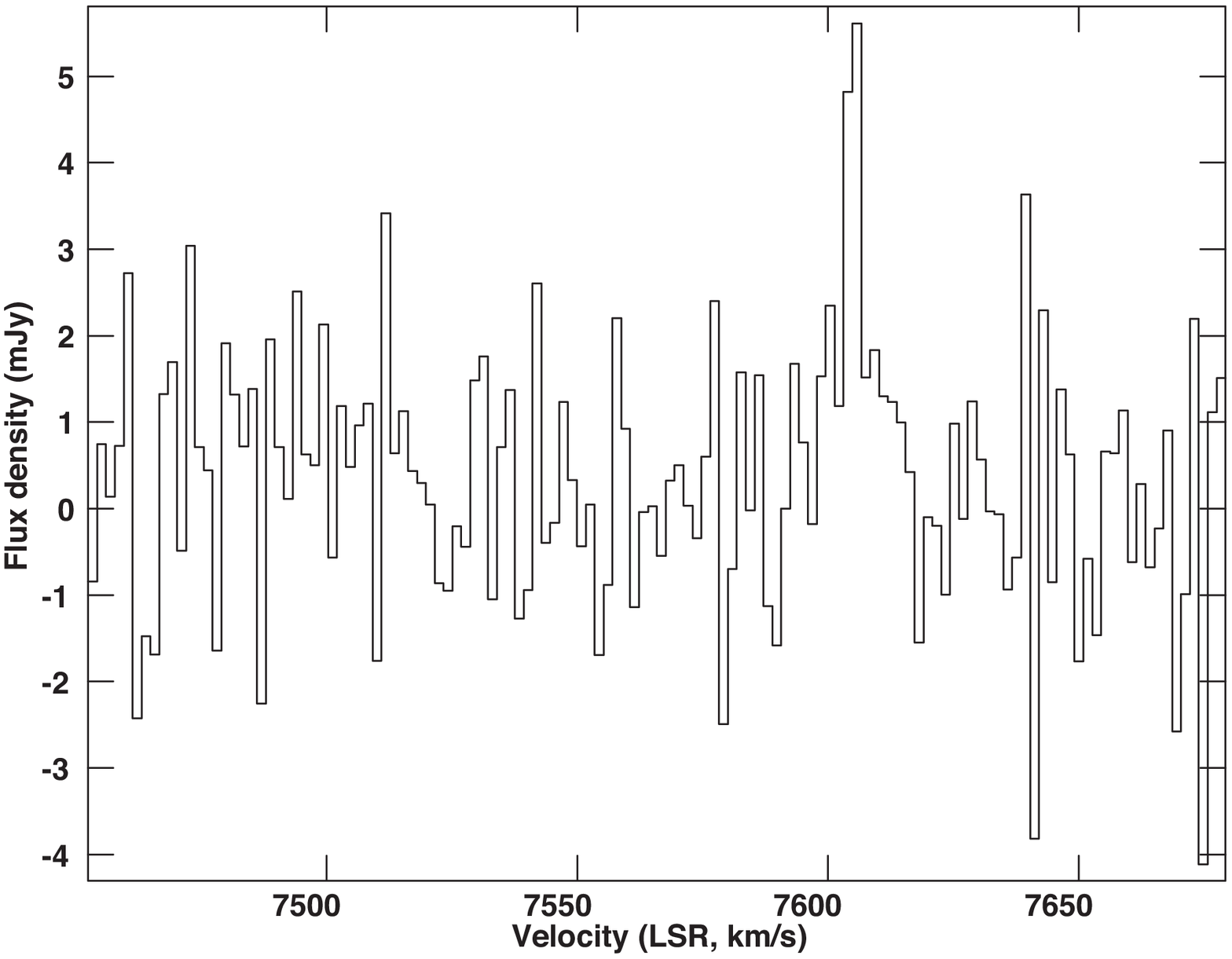}
\caption{Comparison of the 22 GHz \ho maser emission in the southern
  nucleus from 2-epoch VLA observations from the same velocity
  range. Left: the first epoch spectrum, observed in 2012 July. The
  maser features peaked at \vlsr=7563.2 \kms and \vlsr=7609.3 \kms
  (Gaussian-fitted) were seen.  Right: the second epoch spectrum,
  obtained in 2012 August. The \vlsr=7609.3 \kms was seen, while the
  \vlsr=7563.2 \kms feature faded away in this epoch.
 \label{figure1}}
\end{figure}
%
\begin{figure}
\epsscale{1.0}\plottwo{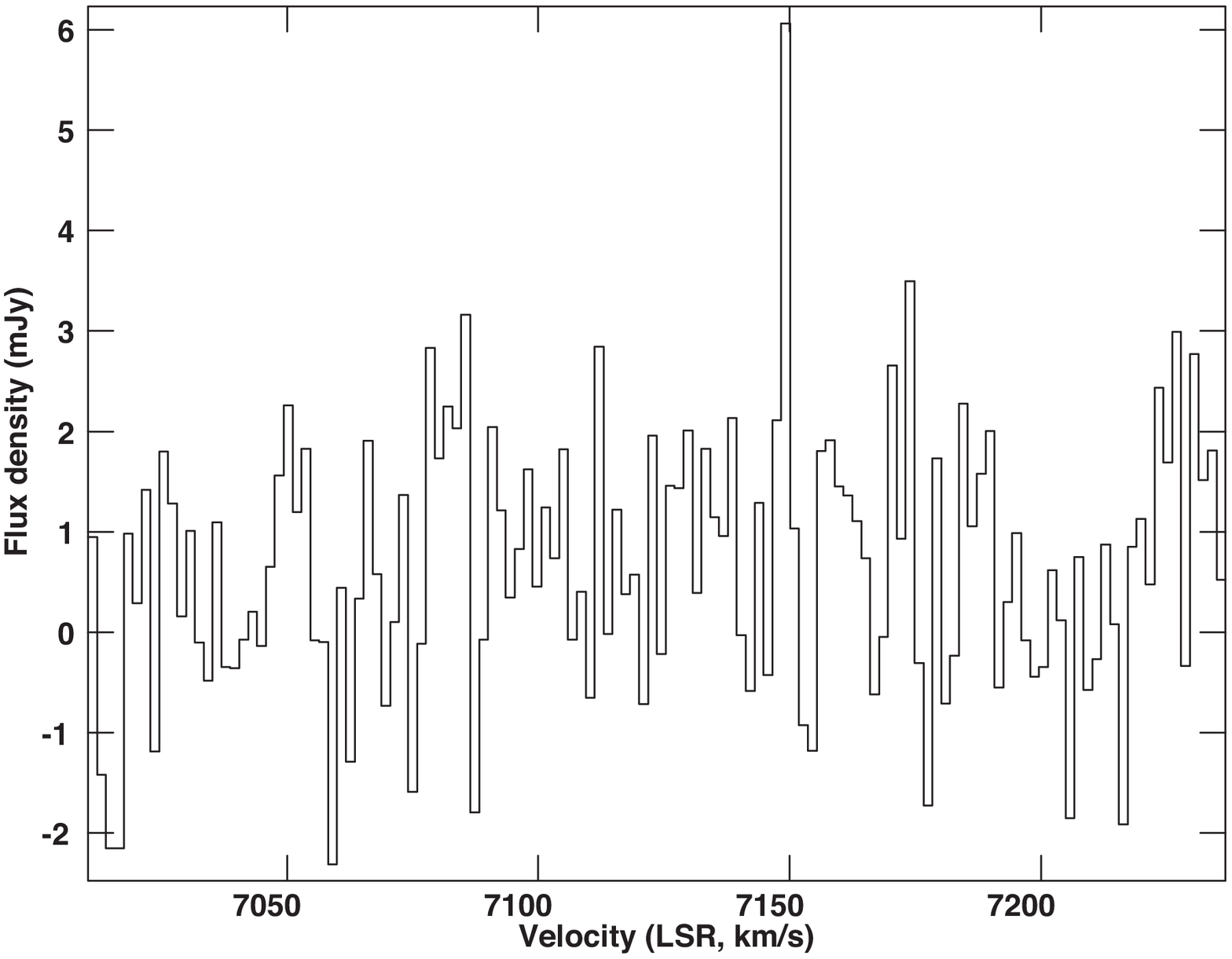}{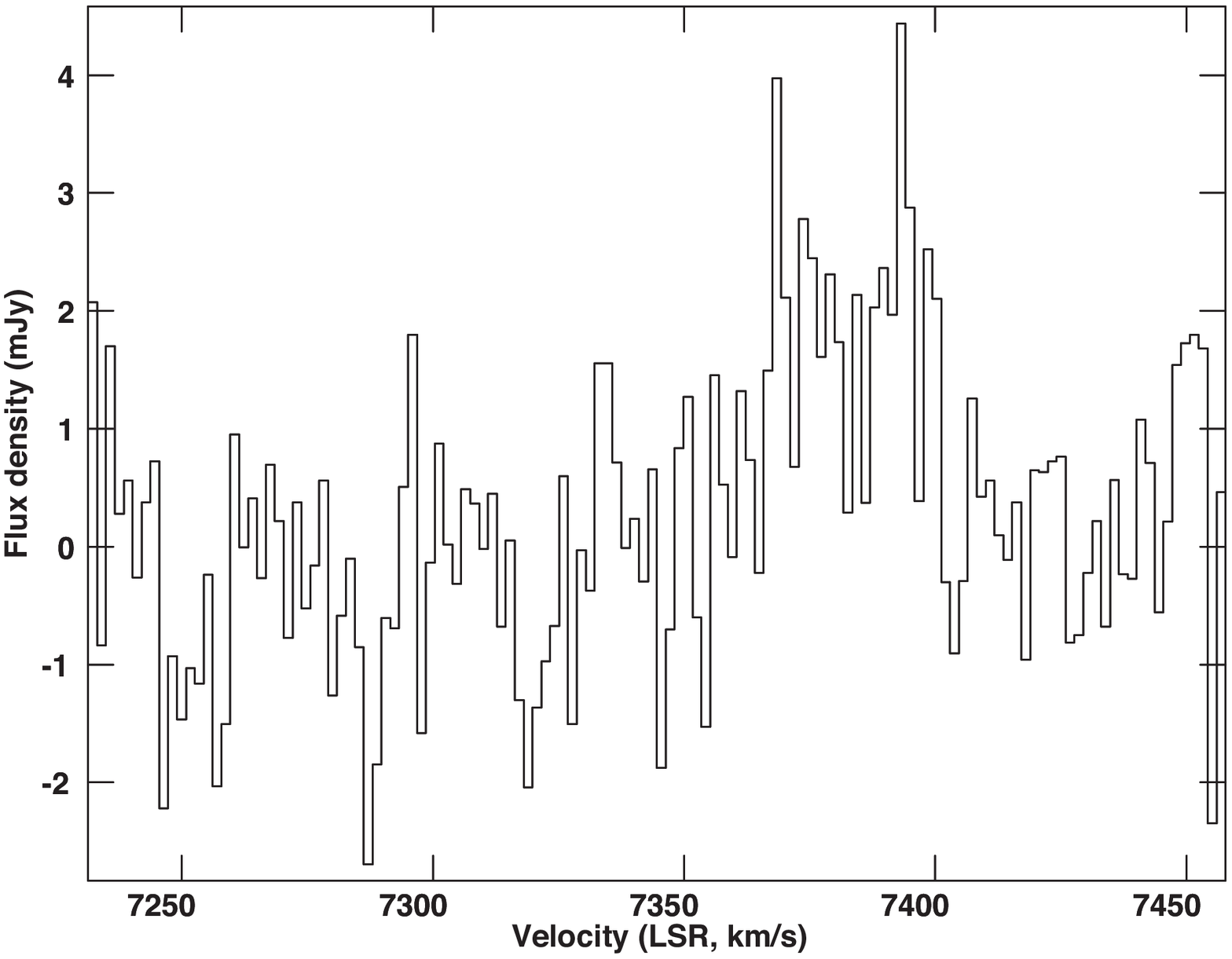}
\caption{{ Tentative detection of the 22 GHz \ho maser that consists of 
 narrow features peaked
(left) at \vlsr=7158 \kmss, (right) at \vlsr=7372 \kmss, \vlsr=7396 \kmss, 
   and a broad feature (\vlsr $\approx$ 7360 —- 7400 \kmss) toward the southern nucleus}, all of which were obtained in
  the second epoch. {No features are detected at the systemic velocity (7304 \kmss).} The peak velocities were Gaussian-fitted values.
  The detection needs to be confirmed.\label{figure2}}
\end{figure}

\begin{figure}
\epsscale{0.5}\plotone{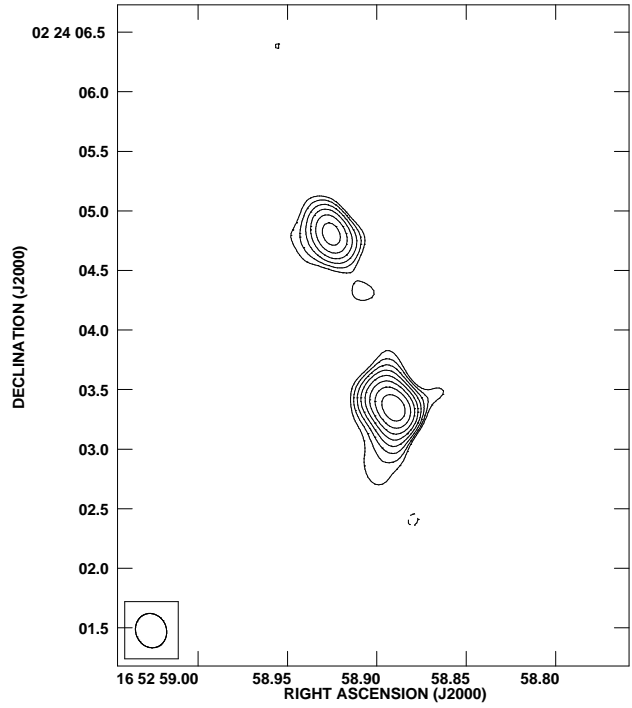}
\caption{22 GHz continuum image of NGC 6240 (256 MHz
  bandwidth), obtained from the single-epoch VLA observation in 2012 July. 
The synthesized beam (FWHM) is plotted in the bottom left 
  corner of the image. The contour levels
  are --4, 4, 5.6, 8, 11.3, 16, 22.6, 32, 45.2 of one $\sigma$ value of 0.13 \mb and the peak flux
  density is 7.75 \mb. 
\label{figure3}}
\end{figure}


\begin{figure}
\epsscale{0.6}\plotone{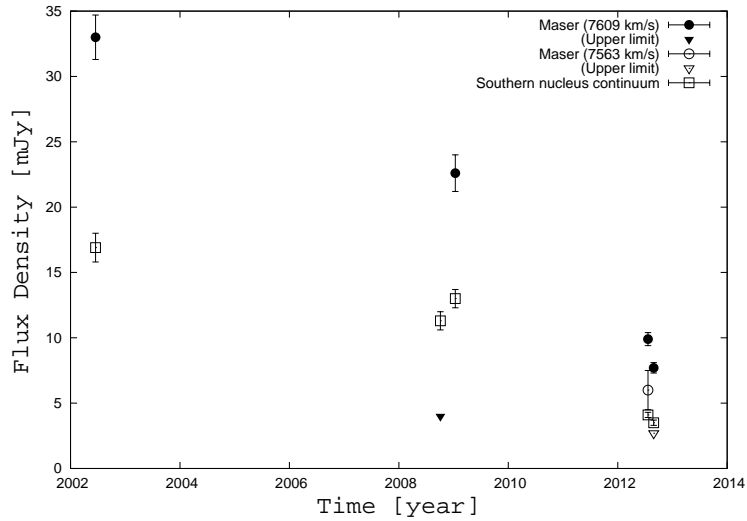}
\caption{Flux densities of the \ho masers (filled and open circles) and 22 GHz continuum of the southern nucleus (open square), measured by the VLA in two epochs in 2012, are presented.
\label{figure4}}
\end{figure}
\begin{figure}
\epsscale{1.0}\plottwo{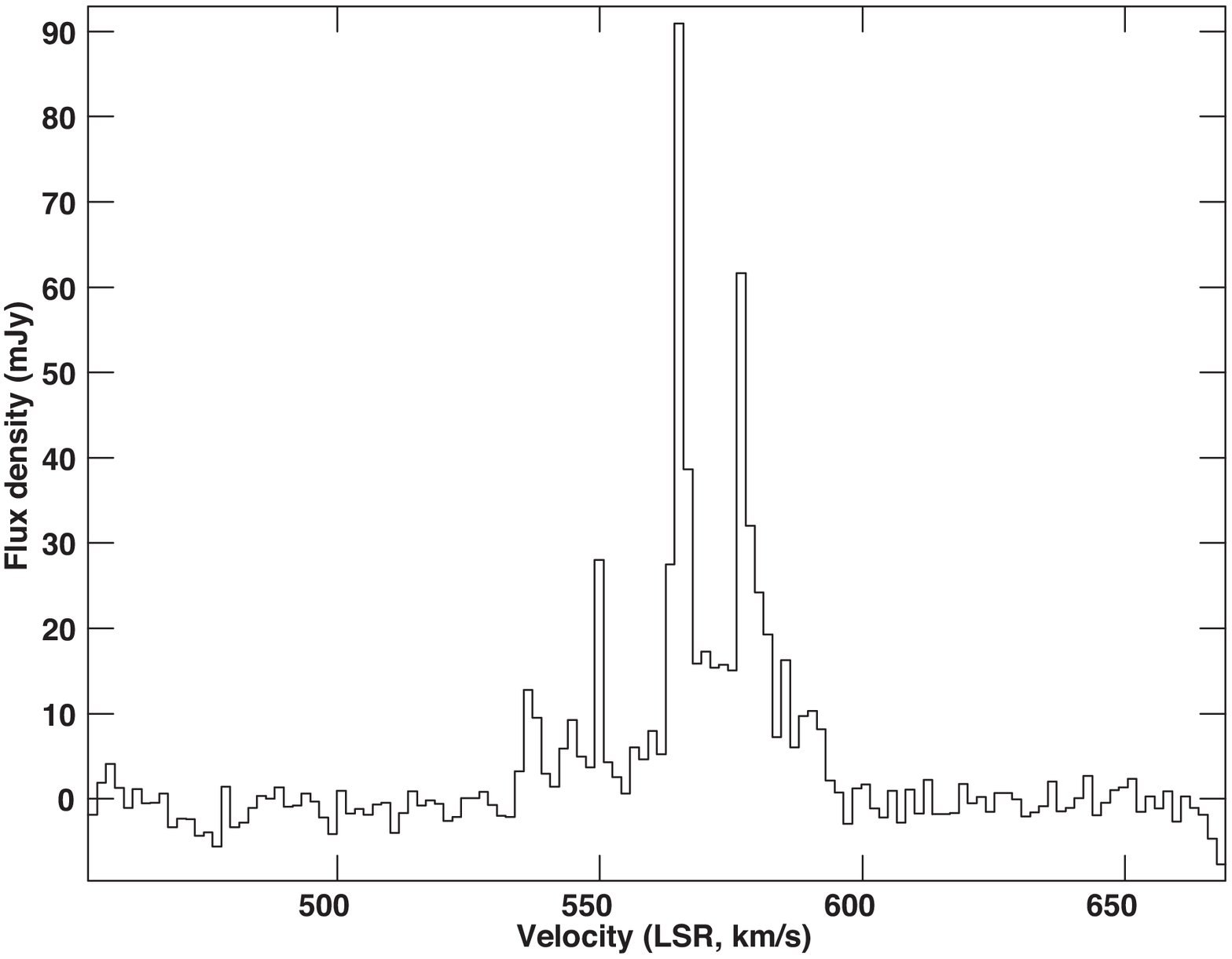}{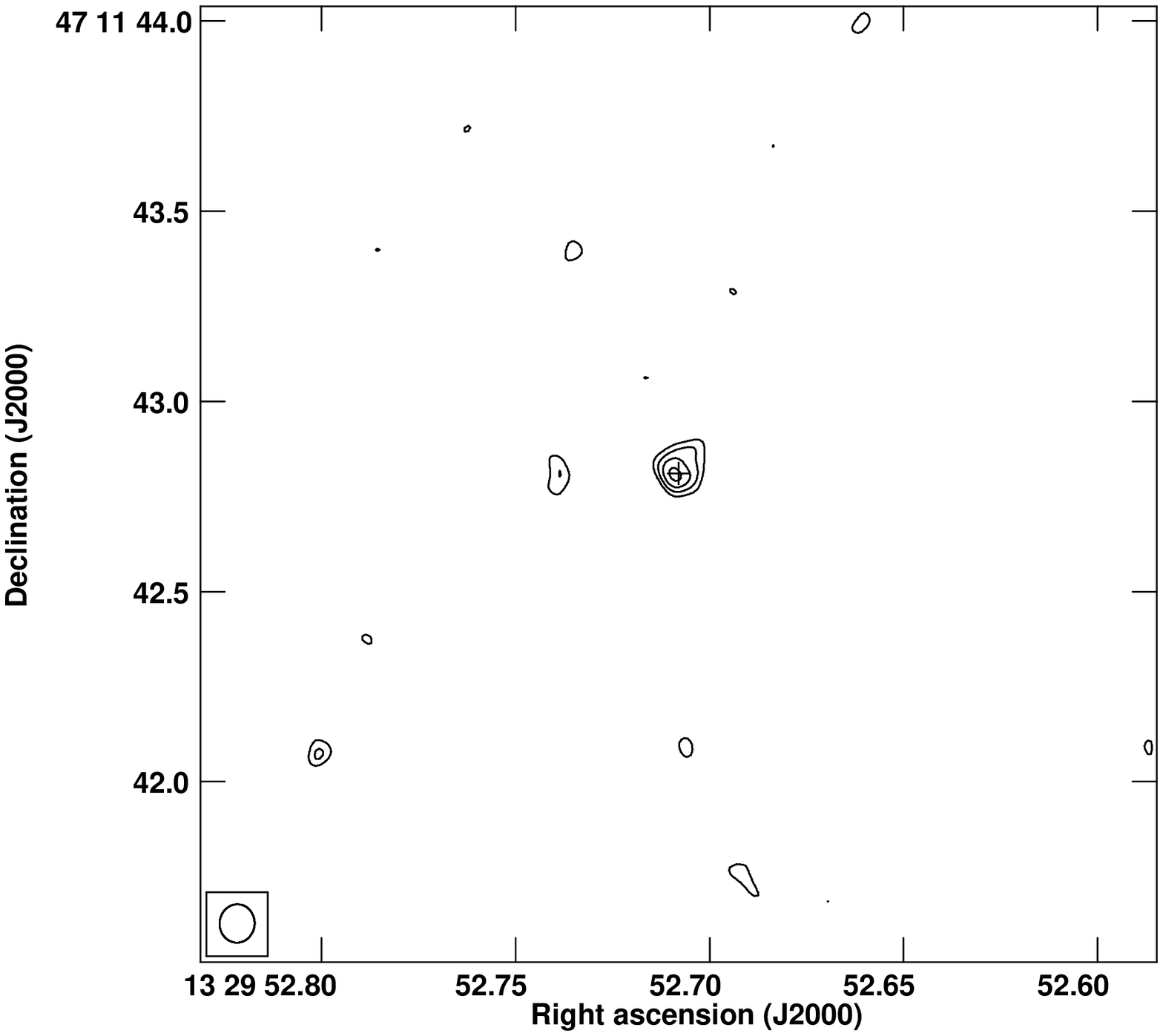}
\caption{\ho maser spectrum and continuum image of M51, observed in
  2012 November, by the VLA in the A configuration are presented.
  Left: 22 GHz \ho maser spectrum between \vlsr = 460-- 670 \kms
  toward the center of the galaxy.  
  Right: 22 GHz continuum image of
  the galaxy obtained by using 256 MHz bandwidth.  The contour levels
  are 3,4,5,6 of one $\sigma$ value of 0.0365 \mb~ and the peak flux
  density is 0.231 \mb. The position of the detected \ho maser (\vlsr=549, 563, 565, 576\kms 
  and some other minor features) is marked by a cross. \label{figure5}}
\end{figure}
\end{onecolumn}

\end{document}